\documentclass[aps,pre,twocolumn,showpacs,nobalancelastpage, floatfix]{revtex4-1}

\usepackage{epsfig,amssymb,amsmath,graphicx,subfigure,hyperref}
\usepackage{tabularx}
\usepackage{float}
\usepackage{setspace}
\usepackage{color}
\usepackage{times}
\usepackage{verbatim}

\begin{document}
\title{Odd moduli of disordered odd elastic lattices}
\author{Zhitao Chen}
\affiliation{Department of Physics, University of California Santa Barbara, Santa Barbara, California 93106, USA}
\email{E-mail: zhitao_chen@ucsb.edu}

\begin{abstract}
We study the effects of bond disorder on triangular and honeycomb lattices where each spring has a probability $p$ to be odd elastic. Using an effective medium theory and numerical simulations, we uncover the behavior of odd moduli in the presence of disorder, which we interpret as a crossover between the affine response of the passive elastic backbone, and a rigidity percolation transition in the odd elastic components. Though oddness is generally robust against disorder even at low $p$, we find that fine-tuned features of an odd elastic honeycomb lattice are not robust against disorder. 
\end{abstract}

\maketitle

\section{Introduction}
An odd elastic material is an elastic system whose dynamics is not governed by an elastic potential energy \cite{Scheibner2020}. In the Hookean approximation, its stress and strain are related by a non-symmetric elastic tensor. 
The absence of a potential energy gives rise to peculiar properties in odd elastic systems, including work extraction from a deformation cycle, self-sustained waves in the bulk \cite{Scheibner2020}, and non-Hermitian skin effects of topological origin \cite{Zhou2020,ScheibnerPRL}. 
The presence of nonreciprocity due to the non-symmetric elastic tensor makes odd elastic material an example of a broad class of systems known as active matter \cite{Active20}.  
Experimentally, odd elasticity has been achieved in a robotics system with piezoelectric elements \cite{Chen2021}, in a colloidal spinners system \cite{Bililign2022}, and has been observed in assemblies of starfish embryos which exhibit self-sustained chiral waves \cite{Tan2021}.

Given the potential of work extraction from engines made of odd elastic material, it is natural to ask what are the effects of disorder on an odd elastic medium. How do the odd moduli behave if some electronic parts in an artificial odd system malfunction? How robust is odd elasticity against disorder? Does disorder induce a passive-to-odd phase transition?
These questions may also be relevant in the context of the starfish embryos experiment \cite{Tan2021}. 
As embryos develop, their mutual interaction suffers an increase of effective noise, which leads to an eventual dissolution of the odd crystal.
Studying disorder may shed light on the properties of the odd crystal in the intermediate to long time frame.

Here we study the effects of bond disorder on the odd moduli of two dimensional triangular and honeycomb lattices, where each spring has a probability $p$ of being odd. 
We find that odd elasticity is robust against disorder, and it persists at small $p$, i.e. there is no passive-to-odd phase transition at a finite $p$. 
The behavior of odd moduli as a function of $p$ is a result of a competition between the affine response of the passive elastic backbone, and a rigidity percolation transition in the odd elastic components.
We also see that a fine-tuned odd elastic modulus in the honeycomb lattices is not stable against disorder.
Note that in this work, we refer to elasticity that is not odd as \textit{passive} since odd elasticity is an example of active matter, and implementing odd elasticity often requires external sources of energy. However, a passive realization of odd elasticity has been recently proposed in \cite{Shaat2021}.

\section{Disordered odd elastic lattices}
For triangular lattices, we impose that each bond is at least a passive spring with a spring constant $k$. Additionally, each spring has a probability $p$ of having an odd spring constant $k^o$. More explicitly, an odd spring with rest length $l$ connecting particles at $\mathbf{r_1}$ and $\mathbf{r_2}$ exerts a force
\begin{equation}
\mathbf{F}=-\left(k \frac{\Delta \mathbf{r}}{|\Delta\mathbf{r}|} +k^o \frac{\Delta \mathbf{r}^*}{|\Delta \mathbf{r}^*|}\right)\left(|\Delta\mathbf{r}|-l\right)
\label{eq:force}
\end{equation}
on the particle at $\mathbf{r_1}$ \cite{Scheibner2020}. Here $\Delta \mathbf{r}=\mathbf{r_1}-\mathbf{r_2}$, and $\Delta r^*_i=\epsilon_{ij} \Delta r_j$, where $\epsilon_{ij}$ is the Levi-Civita symbol.

For honeycomb lattices, every nearest neighbor (NN) and next nearest neighbor (NNN) spring has a passive spring constant $k$. Each NN spring has a probability $p$ of having an odd spring constant $k^o_1$. Likewise, each NNN spring has a probability $p$ of having $k^o_2$.

At the disorder-free $p=1$ limit, both cases are examples of isotropic odd elastic materials. In the continuum limit and to linear order, one can write down a coarse-grained stress strain relation with a non-symmetric elastic tensor (see Ref \cite{Scheibner2020} for details). 
Expanding stress in the ordered basis of pressure ($\sigma^0$), torque density ($\sigma^1$), and shear stresses in different directions ($\sigma^2$ and $\sigma^3$), and similarly expanding strain into its corresponding basis $\{u^i\}$ of dilation, rotation, and the two shear strains, we have the stress strain relation
\begin{equation}
\begin{pmatrix}
\sigma^0\\ \sigma^1\\ \sigma^2 \\ \sigma^3
\end{pmatrix}
=2\begin{pmatrix}
B & 0 & 0 & 0\\
A & 0 & 0 & 0\\
0 & 0 & \mu & K^o\\
0 & 0 & -K^o & \mu
\end{pmatrix}
\begin{pmatrix}
u^0 \\ u^1 \\ u^2 \\ u^3
\end{pmatrix}.
\label{eq:stress_strain}
\end{equation}
Here, $B$ is the passive bulk modulus, and $\mu$ is the shear modulus, both proportional to $k$. The off diagonal matrix elements arise due to oddness.
The modulus $A$ couples dilation or compression to a torque density, and $K^o$ couples shears in different directions.
For a triangular lattice, $A=2K^o=\sqrt{3}k^o/2$.
For a honeycomb lattice,
$A=(k^o_1+6 k^o_2)/(2\sqrt{3})$, and $K^o=\sqrt{3} k^o_2 /2$.
One can therefore fine-tune the value of $A$ independent of $K^o$ by choosing $k^o_1$, and we fix $k^o_1/k^o_2 = -6$ so that $A=0$ (torque-free) at $p=1$.
The aim of our study is to determine how these moduli change as functions of $p$, and we will use $A_m$ and $K^o_m$ for the measured odd elastic moduli of the disordered lattices.

\section{Overview of Theoretical and simulations methods}
We study the disordered lattices with a well-established effective medium theory (EMT) with the coherent potential approximation. 
It is a mean field method where we describe the disordered system with a disorder-free effective medium whose parameters are determined self-consistently \cite{Elliott_RMP,Feng1985,Mao2013}.
Here we use the triangular lattice as an example to describe the main idea of EMT.
The linear dynamics of the effective medium is described by a dynamical matrix $\mathbf{D}_m$, and its Green's function $\mathbf{G}_m=-(\mathbf{D}_m)^{-1}$.
The dynamical matrix has the same form as that of a disorder-free system, except that it has an effective odd spring constant $k^o_m$, which is generally not equal to $k^o$ except at $p=1$.
We determine $k^o_m$ in the following way:
We replace a single spring in the effective medium by a disordered spring with an odd elastic constant $k^o_s$.
Mirroring the setup of the original problem, $k^o_s$ is a random variable with a probability distribution
\begin{equation}
P(k^o_s)=p\,\delta(k^o_s - k^o) + (1-p)\,\delta(k^o_s).
\label{eq:prob}
\end{equation}
We describe this replacement with a perturbation $\mathbf{V}$ to the dynamical matrix. The perturbed Green's function of the effective medium then takes the form
\begin{equation}
\mathbf{G}=\mathbf{G}_m+\mathbf{G}_m\cdot \mathbf{T}\cdot \mathbf{G}_m,
\end{equation}
where
\begin{equation}
\mathbf{T}=\mathbf{V}\cdot (\mathbf{I}-\mathbf{G}_m \cdot \mathbf{V})^{-1}.
\label{eq:T}
\end{equation}
Since the effective medium is a model of the original disordered system, we require that $\langle\mathbf{G}\rangle=\mathbf{G}_m$, which gives the self-consistent EMT equation
\begin{equation}
\langle \mathbf{T}\rangle=0,
\label{eq:EMT}
\end{equation}
where the average is taken over the probability distribution in Equation (\ref{eq:prob}).
This gives $k^o_m$ as a function of $p$, $k^o$ and the lattice geometry.
This form of EMT has been successfully applied to problems in various contexts \cite{Soven1969,LiartePRL,Shimada2022}, and it can be adapted naturally to systems without a notion of energy.

We also study the problem with molecular dynamics simulations.
For both triangular and honeycomb lattices, we setup about a thousand unit cells where the springs obey the force law in Equation (\ref{eq:force}) and disorder according to the description in the previous section.
To measure the elastic moduli, we first impose a small global compression or shear with an affine displacement $\mathbf{u}^{\text{aff}}_i=\mathbf{\eta}\mathbf{r}_i$. Here, the $i$ subscript indexes the $N$ vertices in the lattice, and $\mathbf{\eta}$ is a $2$ by $2$ matrix.
For compression, $\mathbf{\eta}=-\gamma \mathbf{I}$, and for shear, $\mathbf{\eta}$ has zeros on the diagonal and $\gamma$ on the off-diagonal. 
In all simulations, we use $\gamma=0.01$.
While keeping the boundary fixed after the initial distortion, we time evolve the bulk of the system in the over-damped limit using a second order Runge-Kutta method. 
We then measure and perform a spatial average of the stress tensor in the bulk, and extract the elastic moduli from Equation (\ref{eq:stress_strain}). We measure all elastic moduli in units of the passive spring constant $k$, and for the rest of this work we omit writing $k$ explicitly.
We also record a non-affine parameter $\Gamma$ for the final configuration of each simulation run, defined as
\begin{equation}
\Gamma=\frac{1}{N\gamma^2}\sum_i \lVert \mathbf{u}_i - \mathbf{u}^{\text{aff}}_i\rVert^2,
\end{equation}
which quantifies how far the final configuration is from the initially imposed distortion \cite{Liarte_2016}. We perform ten independent simulation runs for each choice of parameter.
Since the results of EMT and molecular dynamics simulations agree reasonably well, we present their results collectively below.

\section{Results}
\subsection{Triangular lattices}
For triangular lattices, the EMT equation (\ref{eq:EMT}) gives (see Appendix for more details)
\begin{equation}
k^o_m=k^o\,\frac{p-H(k^o_m)}{1-H(k^o_m)},
\label{eq:tri}
\end{equation}
where $H(k^o_m)=2(k^o_m)^2/\left[3+3(k^o_m)^2\right]$.
After some arithmetics, we see that Equation (\ref{eq:tri}) is a cubic equation with a unique solution $k^o_m$ for any probability $p>0$. This implies that even the slightest presence of odd springs in our setup makes the system odd as a whole. 
This is a natural feature of the design since an odd spring breaks detailed balance locally with a chiral force, and every odd spring in the setup has the same sign of $k^o$, and hence the same sign of chirality. 
The latter is important as one can imagine that in a setup where each odd spring has an equal probability to be $k^o$ and $-k^o$, the system as a whole can have no oddness on average.

\begin{figure}
\centering
\includegraphics[width=3.375 in]{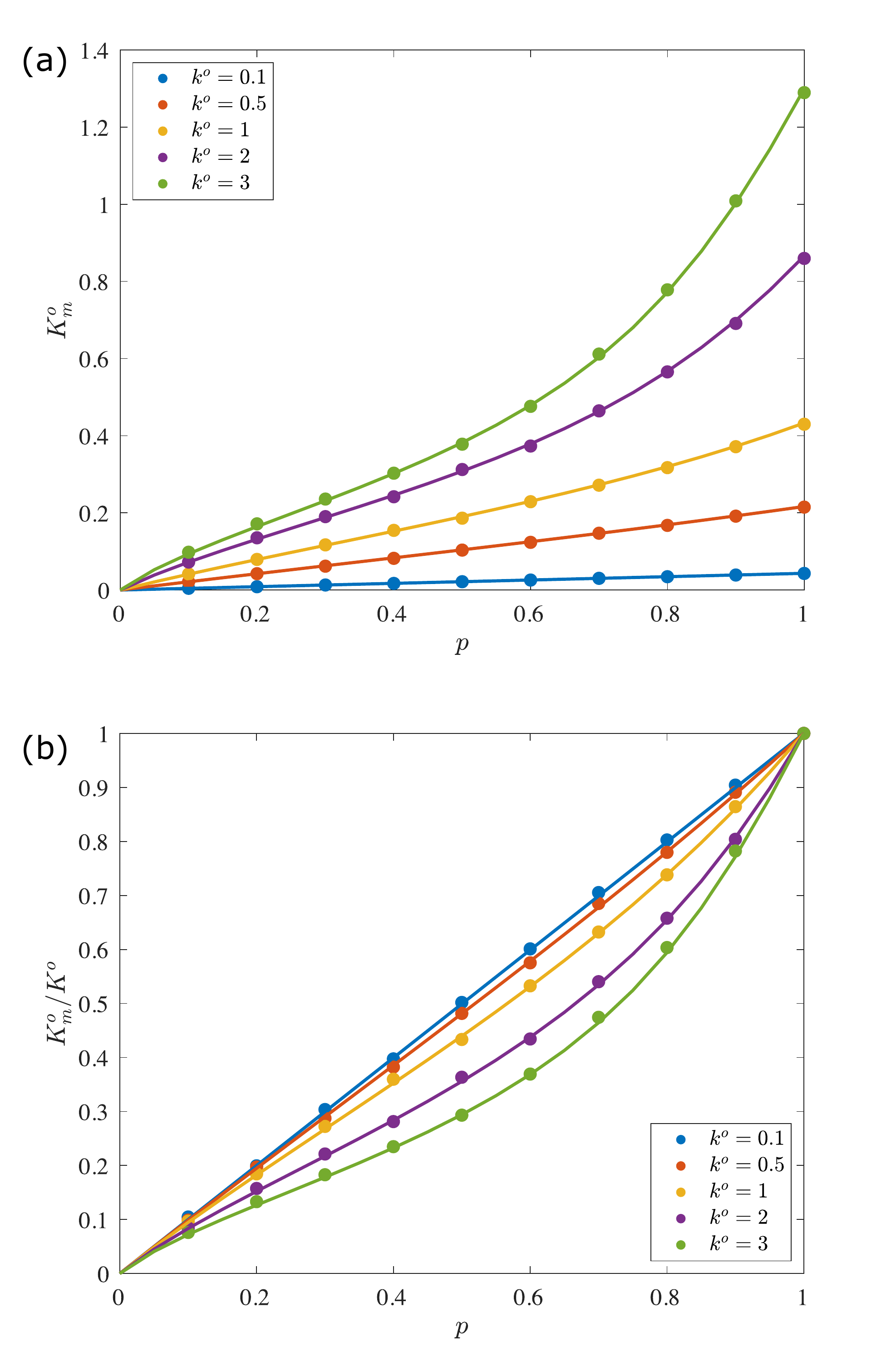}
\caption{Effective odd modulus $K^o_m$, unscaled (a) and scaled (b) as a function of probability $p$, for triangular lattices at different values of odd spring constant $k^o$. Numerical solutions of Equation (\ref{eq:tri}) are plotted with solid curves, and molecular dynamics simulations results are presented with dots. Error bars are smaller than dot size.}
\label{fig:tri}
\end{figure}

We now examine $K^o_m=A_m/2=\sqrt{3} k^o_m/4$ as a function of $p$ and the odd spring constant $k^o$.
At low $p$, we expect that $k^o_m\ll1$. Solving Equation (\ref{eq:tri}) in this regime gives
\begin{equation}
\frac{K^o_m}{K^o}=\frac{k^o_m}{k^o} \approx p-\frac{2}{3}(k^o)^2 p^2.
\end{equation}
That is, after an initial linear ramp at small $p$, the scaled effective oddness increases at a rate negatively correlated with $k^o$ at slightly higher $p$. 
At $p=2/3$, we expect the odd springs in the disordered lattice to form a system spanning cluster, and Equation (\ref{eq:tri}) gives $k^o_m/k^o\lessapprox2/3=p$ for small $k^o$, and $k^o_m/k^o\sim(k^o)^{-5/3}$ for large $k^o$. 
Lastly, near the disorder-free point, expanding with $\delta p=1-p$ and $\delta K^0_m=K^o-K^o_m$, we get
\begin{equation}
\delta K^0_m\approx K^o\, \frac{3(k^o)^2+3}{(k^o)^2 +3}\,\delta p,
\end{equation}
where the fraction on the right hand side is an increasing function of $k^o$.
From these simple calculations, we come to the qualitative conclusion that for small $k^o$, the scaled effective oddness $K^o_m/K^o$ increases almost linearly with $p$. In contrast, for a larger $k^o$, $K^o_m/K^o$ increases slowly at small $p$, and then increases much faster after some probability around $p=2/3$. These features are consistent with molecular dynamics simulations results and numerical solution of Equation (\ref{eq:tri}), presented in Fig.(\ref{fig:tri}).

\begin{figure}
\centering
\includegraphics[width=3.375in]{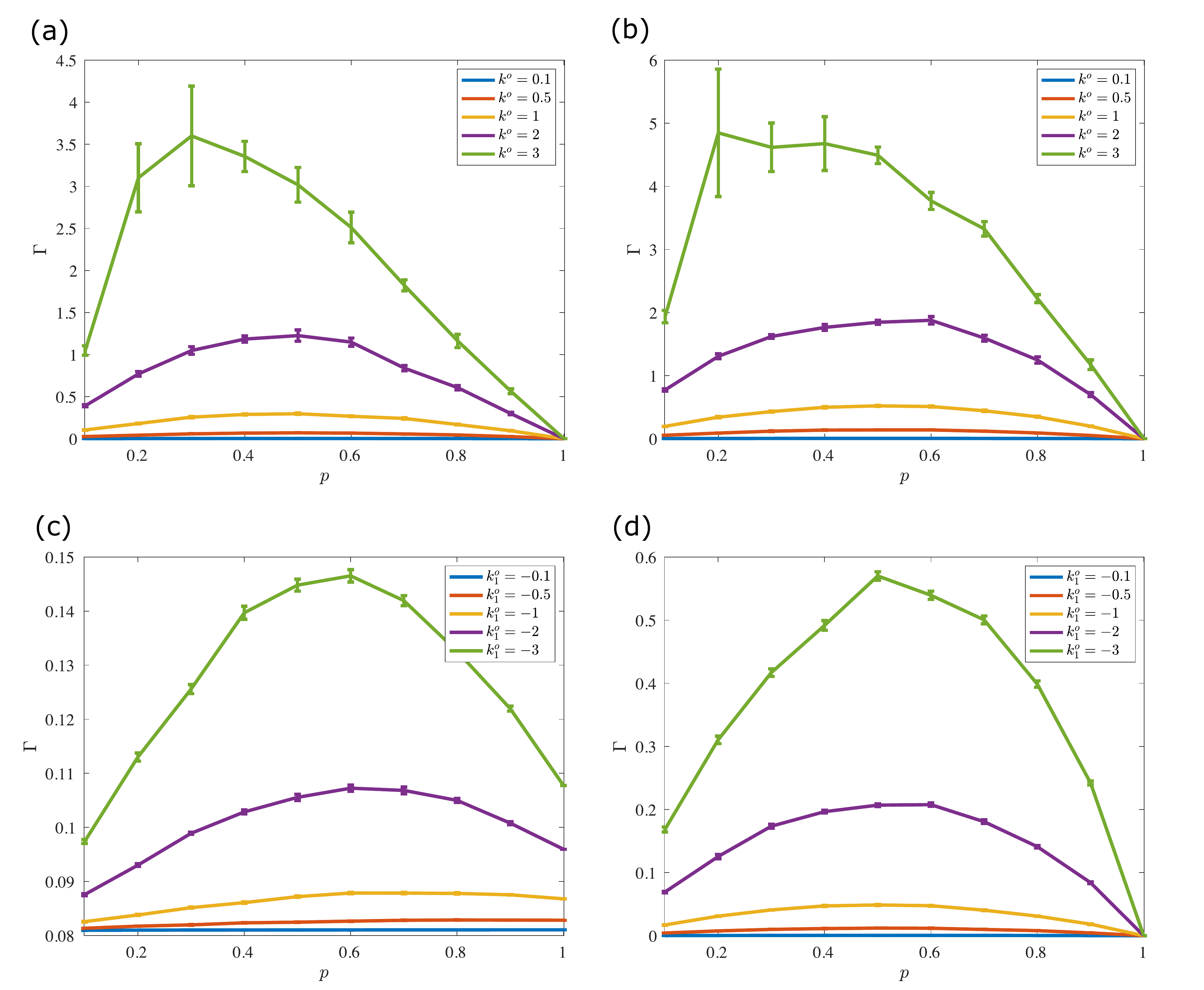}
\caption{Non-affine parameter $\Gamma$ for triangular lattices (top row) and honeycomb lattices (bottom row) with an initial global shear (left column) and compression (right column) distortion.}
\label{fig:Gamma}
\end{figure}

We understand the results as follows: 
In the limit of small $k^o$, the elastic property of the disordered lattice is controlled mostly by the passive elastic backbone, which responds to a small global distortion affinely, giving a linear relation between $K^o_m$ and $p$.
In the extremely large $k^o$ limit, the response of the system is controlled by the odd components, which undergo a rigidity percolation transition at $p=2/3$.
Since in this limit, there is no elastic response below $p=2/3$, we expect $K^o_m$ to be zero for $p<2/3$ before ramping up for $p>2/3$. The actual behavior of $K^o_m$ is then a crossover between these two limits.
Similar to a passive rigidity percolation transition, the probability $p$ at which $K^o_m/K^o$ deviates the most from linearity is near $p=2/3$, as shown in Fig.(\ref{fig:tri}b). 
However, this does not imply that the non-affine parameter $\Gamma$ of the system is the highest at this value.
Indeed, as shown in Fig.(\ref{fig:Gamma}a, b), the non-affine parameter for both sheared and compressed disordered triangular lattices peaks at a lower value of $p$, and is dependent on $k^o$.

\subsection{Honeycomb lattices}
For honeycomb lattices, we have two EMT equations, one for disordered NN springs and the other for disordered NNN springs. 
These equations are
\begin{equation}
\begin{aligned}
k^o_{1,m}=k^o_1\,\frac{p-H_1(k^o_{1,m},k^o_{2,m})}{1-H_1(k^o_{1,m},k^o_{2,m})}\\
k^o_{2,m}=k^o_2\,\frac{p-H_2(k^o_{1,m},k^o_{2,m})}{1-H_2(k^o_{1,m},k^o_{2,m})}.
\end{aligned}
\label{eq:hon}
\end{equation}
The effective odd spring constants are coupled through functions $H_1(k^o_{1,m},k^o_{2,m})$ and $H_2(k^o_{1,m},k^o_{2,m})$.
These are complicated functions with no convenient closed forms, in contrast to the one for triangular lattices, and we present their explicit construction in the Appendix.
Despite such complication, we can expect that in general, $H_1\neq H_2$, and from Equation (\ref{eq:hon}) we see that  $(k^o_{1,m}/k^o_{2,m})\neq (k^o_1/k^o_2)$ for $p<1$.
This means that even though we choose $k^o_1/k^o_2=-6$, making our system torque-free with $A=(k^o_1+6 k^o_2)/(2\sqrt{3})=0$ at $p=1$, this feature and, in general, any fine-tuned ratio between $A$ and $K^o$, are not robust in the presence of disorder. 

\begin{figure}
\centering
\includegraphics[width=3.375 in]{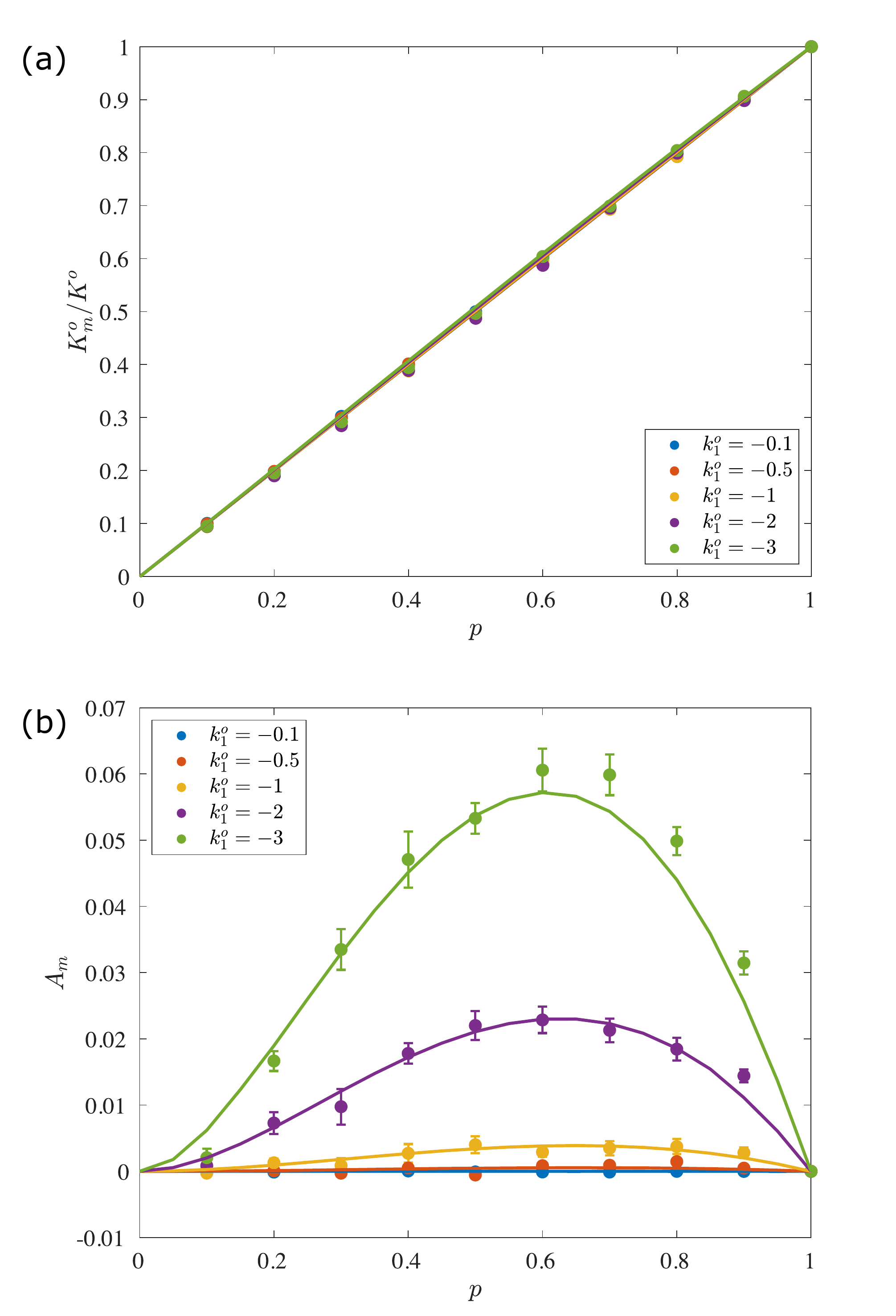}
\caption{Effective odd modulus $K^o_m$ (a) and $A_m$ (b) in disordered odd honeycomb lattices, at different values of $k^o_1$, with a fixed ratio $k^o_1/k^o_2=-6$. Numerical solutions of Equation (\ref{eq:hon}) are plotted with solid curves, and molecular dynamics simulations results are presented with dots. The fine-tuned torque-free feature ($A=0$) is broken by disorder. }
\label{fig:hon}
\end{figure}

We solve Equation (\ref{eq:hon}) numerically and the results compare well with molecular dynamics simulations, as shown in Fig.(\ref{fig:hon}).
We see that the odd modulus $K^o_m$ is close to linear in $p$, and as expected, $A_m$ is non-zero in general except at $p=1$, peaking near $p=0.6$.
We offer an understanding of these results as follows, building upon the intuition from studying disordred triangular lattices.
First, recall that $K^o_m=\sqrt{3}\,k^o_{2,m}/2$ is completely determined by the NNN springs, which form two overlapping triangular lattices. 
Since we use $k^o_2\leq 1/2$, it is not surprising that  $K^o_m$ grows almost linearly as a function of $p$ for all our simulations.
As in the triangular lattices, for small (absolute) values of $k^o_1$, we also expect $k^o_{1,m}$ to be almost linear in $p$, making $A\approx0$, as shown in Fig.(\ref{fig:hon}a).
At a larger value of $k^o_1$, we expect $k^o_{1,m}$ to deviate from this linear behavior, satisfying $k^o_{1,m}(p)/k^o_1<p$, as suggested by Fig.(\ref{fig:tri}b).
Combining the behaviors of $k^o_{1,m}$ and $k^o_{2,m}$, we see that $A_m$ should be positive for $0<p<1$, peaking at a value of $p$ at which $k^o_{1,m}$ deviates the most from linearity.
Therefore, even though $k^o_{1,m}$ and $k^o_{2,m}$ are coupled in the EMT equations, we can understand them qualitatively by considering them independently.
Finally, we observe that the non-affine parameter $\Gamma$ in a honeycomb lattice has similar qualitative features as it does in a triangular lattice, as shown in Fig.(\ref{fig:Gamma}). At low $p$, it increases as a function of $p$ due to the presence of disorder and odd elasticity. It eventually drops down to zero (or a smaller value) in the disorder-free limit for an initially compressed (sheared) honeycomb lattice.

In conclusion, we have studied the effective odd moduli in disordered triangular and honeycomb lattices. A possible future direction is to study the dynamical behavior of disordered odd elastic networks. Perhaps more physically and relevant to experiments \cite{Tan2021,Bililign2022}, one can study passive-odd elastic mixture without fixed connectivity.

\begin{acknowledgements}
I thank Mark Bowick and Cristina Marchetti for encouragement and support. I thank Austin Hopkins for introducing me to Numba, which accelerated this project by over an order of magnitude, and for reading the first draft of this work.
\end{acknowledgements}

\bibliography{refs}

\appendix*
\section{Matrices and functions in EMT}
For completeness, we present here some of the mathematical expressions in EMT with more details.
In Hookean mechanics, the dynamical matrix in real space is defined as
\begin{equation}
\mathbf{F}_l=-\sum_{l'}\mathbf{D}_{ll'}\mathbf{u}_{l'},
\end{equation}
where $\mathbf{F}_l$ denotes forces on particles in unit cell $l$, and $\mathbf{u}_l'$ denotes the displacements of particles in unit cell $l'$. Here, the subscript $l$ indexes the $N_c$ number of unit cells in the lattice.
$\mathbf{D}_{ll'}$ is $2$ by $2$ for triangular lattices and $4$ by $4$ for honeycomb lattices. 
Since the effective medium in EMT is disorder-free, we choose to work with the dynamical matrix in Fourier space.
In a triangular lattice with passive spring constant $k$ and odd spring constant $k^o$, the dynamical matrix is
\begin{equation}
\mathbf{D}_{\mathbf{q}\mathbf{q'}}=N_c\delta_{\mathbf{q}\mathbf{q'}}\left(k\sum_{n=1}^3 \mathbf{b}_{n,\mathbf{q}}\mathbf{b}_{n,-\mathbf{q}}+k^o\sum_{n=1}^3 \mathbf{b}_{n,\mathbf{q}}^*\mathbf{b}_{n,-\mathbf{q}} \right).
\label{eq:D_tri}
\end{equation}
The vectors are defined as $\mathbf{b}_{n,\mathbf{q}}=\hat{\mathbf{e}}_n [1-\text{exp}(-i\mathbf{q}\cdot\hat{\mathbf{e}}_n)]$ \cite{Mao2013}.
And the unit vectors connecting nearest neighbors on the lattice are $\hat{\mathbf{e}}_1=(1,0)^T$, $\hat{\mathbf{e}}_2=(1/2,\sqrt{3}/2)^T$, and $\hat{\mathbf{e}}_3=(-1/2,\sqrt{3}/2)^T$.
The asterisk superscript denotes a contraction with the Levi-Civita symbol, and adjacency of two vectors denotes an outer product.
In EMT, we replace a single spring in the effective medium by a disordered spring. Without loss of generality, we choose this spring to be at the origin and in the $\hat{\mathbf{e}}_1$ direction.
This introduces a perturbation to the effective medium dynamical matrix in the form of
\begin{equation}
\mathbf{V_{\mathbf{q}\mathbf{q'}}}=(k^o_s - k^o_m)\mathbf{b}_{1,\mathbf{q}}^*\mathbf{b}_{1,-\mathbf{q}},
\end{equation}
where $k^o_s$ is a random variable with a probability distribution given in Equation~(\ref{eq:prob}).
The $T$ matrix from Equation~(\ref{eq:T}) then has the form $T_{\mathbf{q}\mathbf{q'}}=T\,\mathbf{b}_{1,\mathbf{q}}^*\mathbf{b}_{1,-\mathbf{q}}$, with 
\begin{equation}
T=\frac{k^o_s - k^o_m}{1-(k^o_s - k^o_m)\sum_{\mathbf{q}} \frac{1}{N_c}\mathbf{b}_{1,-\mathbf{q}}\cdot\mathbf{G}_{m,\mathbf{q}}\cdot\mathbf{b}^*_{1,\mathbf{q}}},
\end{equation}
where $(-\mathbf{D_{m,\mathbf{q}\mathbf{q'}}})^{-1}=(1/N_c)\delta_{\mathbf{q}\mathbf{q'}}\mathbf{G_{m,\mathbf{q}}}$.
$H(k^o_m,k)$ is defined as
\begin{equation}
H(k^o_m,k)\equiv-k^o_m\,\sum_{\mathbf{q}} \frac{1}{N_c}\mathbf{b}_{1,-\mathbf{q}}\cdot\mathbf{G}_{m,\mathbf{q}}\cdot\mathbf{b}^*_{1,\mathbf{q}}.
\end{equation}
In practice, the sum over the first Brillouin zone is converted into an integral, and the result is $H(k^o_m,k)=2/\left[3(1+(k/k^o_m)^2)\right]$, as given in the main text. The EMT equation is then obtained by requiring $\langle T \rangle=0$.

In a honeycomb lattice with passive spring constant $k$, NN odd spring constant $k^o_1$, and NNN odd spring constant $k^o_2$, the dynamical matrix is given by
\begin{equation}
\mathbf{D}_{\mathbf{q}\mathbf{q'}}=N_c\delta_{\mathbf{q}\mathbf{q'}}(\mathbf{D}_{1,\mathbf{q}}+\mathbf{D}_{2,\mathbf{q}}),
\label{eq:D_hon}
\end{equation}
Here $\mathbf{D}_2$ captures the effect of NNN springs, which form two overlapping triangular lattices. Hence $\mathbf{D}_2$ can be deduced from the dynamical matrix of a triangular lattice. Defining $\mathbf{c}_{n,\mathbf{q}}=(\mathbf{b}_{n,\mathbf{q}},\mathbf{0})^T$, and $\mathbf{d}_{n,\mathbf{q}}=(\mathbf{0}, \mathbf{b}_{n,\mathbf{q}})^T$, we can write down the $4$ by $4$ matrix
\begin{equation}
\begin{aligned}
\mathbf{D}_{2,\mathbf{q}}=&\left(k\sum_{n=1}^3 \mathbf{c}_{n,\mathbf{q}}\mathbf{c}_{n,-\mathbf{q}}+k^o_2\sum_{n=1}^3 \mathbf{c}_{n,\mathbf{q}}^*\mathbf{c}_{n,-\mathbf{q}} \right)\\
+&\left(k\sum_{n=1}^3 \mathbf{d}_{n,\mathbf{q}}\mathbf{d}_{n,-\mathbf{q}}+k^o_2\sum_{n=1}^3 \mathbf{d}_{n,\mathbf{q}}^*\mathbf{d}_{n,-\mathbf{q}} \right).
\end{aligned}
\end{equation}
$\mathbf{D}_1$ captures the effect of NN springs.
We introduce the following sets of vectors:
$\hat{\mathbf{a}}_1=(0,1)^T$, $\hat{\mathbf{a}}_2=(-\sqrt{3}/2,-1/2)^T$, $\hat{\mathbf{a}}_3=(\sqrt{3}/2,-1/2)^T$.
$\mathbf{f}_1=\mathbf{0}$, $\mathbf{f}_2=\hat{\mathbf{e}}_2$, and $\mathbf{f}_3=\hat{\mathbf{e}}_3$, where $\hat{\mathbf{e}}_i$ are the same as before.
Lastly, we define
$\mathbf{g}_{n,\mathbf{q}}=(-\hat{\mathbf{a}}_n, \hat{\mathbf{a}}_n\text{exp}(i\mathbf{q}\cdot\mathbf{f}_n))^T$ and $\mathbf{g}_{n,\mathbf{q}}^*=(-\hat{\mathbf{a}}_n^*, \hat{\mathbf{a}}_n^*\text{exp}(i\mathbf{q}\cdot\mathbf{f}_n))^T$, similar to Ref. \cite{LiartePRL}. Then we have the $4$ by $4$ matrix
\begin{equation}
\mathbf{D}_{1,\mathbf{q}}=k\sum_{n=1}^3 \mathbf{g}_{n,\mathbf{q}}\mathbf{g}_{n,-\mathbf{q}}+k^o_1\sum_{n=1}^3 \mathbf{g}_{n,\mathbf{q}}^*\mathbf{g}_{n,-\mathbf{q}}.
\end{equation}

Similar to the triangular case, we introduce perturbations
\begin{equation}
\begin{aligned}
\mathbf{V_{1,\mathbf{q}\mathbf{q'}}}=(k^o_{1,s} - k^o_{1,m})\mathbf{g}_{1,\mathbf{q}}^*\mathbf{g}_{1,-\mathbf{q}}\\
\mathbf{V_{2,\mathbf{q}\mathbf{q'}}}=(k^o_{2,s} - k^o_{2,m})\mathbf{c}_{1,\mathbf{q}}^*\mathbf{c}_{1,-\mathbf{q}}.
\end{aligned}
\end{equation}
to the effective medium.
These lead to the $\mathbf{T}$ matrices, $\mathbf{T_1}=T_1\mathbf{g}_{1,\mathbf{q}}^*\mathbf{g}_{1,\mathbf{-q}}$, and $\mathbf{T_2}=T_1\mathbf{c}_{1,\mathbf{q}}^*\mathbf{c}_{1,\mathbf{-q}}$, where 
\begin{equation}
T_1=\frac{k^o_{1,s} - k^o_{1,m}}{1-(k^o_{1,s} - k^o_{1,m})\sum_{\mathbf{q}} \frac{1}{N_c}\mathbf{g}_{1,-\mathbf{q}}\cdot\mathbf{G}_{m,\mathbf{q}}\cdot\mathbf{g}^*_{1,\mathbf{q}}},
\end{equation}
and
\begin{equation}
T_2=\frac{k^o_{2,s} - k^o_{2,m}}{1-(k^o_{2,s} - k^o_{2,m})\sum_{\mathbf{q}} \frac{1}{N_c}\mathbf{c}_{1,-\mathbf{q}}\cdot\mathbf{G}_{m,\mathbf{q}}\cdot\mathbf{c}^*_{1,\mathbf{q}}}.
\end{equation}
Here $\mathbf{G}_{m,\mathbf{q}}$ is the effective medium Green's function obtained by inverting Equation (\ref{eq:D_hon}).
The two functions $H_1$ and $H_2$ in the main text are defined as
\begin{equation}
H_1(k^o_{1,m},k^o_{2,m})=-k^o_{1,m}\,\sum_{\mathbf{q}} \frac{1}{N_c}\mathbf{g}_{1,-\mathbf{q}}\cdot\mathbf{G}_{m,\mathbf{q}}\cdot\mathbf{g}^*_{1,\mathbf{q}},
\end{equation}
and
\begin{equation}
H_2(k^o_{1,m},k^o_{2,m})=-k^o_{2,m}\,\sum_{\mathbf{q}} \frac{1}{N_c}\mathbf{c}_{1,-\mathbf{q}}\cdot\mathbf{G}_{m,\mathbf{q}}\cdot\mathbf{c}^*_{1,\mathbf{q}}.
\end{equation}
Requiring $\langle T_1\rangle=0$ and $\langle T_2\rangle=0$ gives the EMT equations in the main text. In contrast to the case with triangular lattices, we do not find any convenient analytical expression of $H_1$ and $H_2$, and therefore resort entirely to numerically solving the EMT equations.

\end{document}